\documentclass[preprint]{aastex}

\usepackage{epsfig}

\slugcomment{Poster presented at JD 8, IAU, Manchester, August, 2000}

\def\lsim{\lower.5ex\hbox{$\; \buildrel < \over \sim \;$}}
\def\gsim{\lower.5ex\hbox{$\; \buildrel > \over \sim \;$}}

\begin{document}

\title{\huge LiBeB, O and Fe Evolution: Refractory Element and 
Cosmic Ray Transport Delays}

\author{R. Ramaty$^1$, B. Kozlovsky$^2$, R. E. Lingenfelter$^3$}
\affil{$^1$NASA/GSFC, Greenbelt, MD 20771, USA\\
$^2$Tel Aviv University, Israel\\
$^3$University of California, San Diego, USA}

\section{\huge Description}

{\large In two recent publications (Ramaty et al. 2000a; Ramaty, 
Lingenfelter \& Kozlovsky 2000b) we developed an Fe, O and  LiBeB 
evolutionary model based on a Monte-Carlo approach. We allowed each 
core collapse supernova  to eject Fe and O according to the 
calculations of Woosley \& Weaver (1995, hereafter WW95) or the 
values given by Tsujimoto \& Shigeyama (1998, hereafter TS). For Fe, 
we added the contributions of thermonuclear supernovae, as described 
in detail in Ramaty et al. (2000a). We calculated the LiBeB produced 
by each supernova by imparting  10$^{50}$ ergs to cosmic rays with a 
source spectrum identical to that of the current epoch cosmic rays. 
For the cosmic ray source composition we assumed two models: CRS and 
CRI. In the CRS model the composition is constant, independent of 
[Fe/H]. Such a composition could arise (Lingenfelter, Higdon \& 
Ramaty 2000) from cosmic ray acceleration in superbubbles from a 
suprathermal population generated from the sputtering of, and 
scattering by fast supernova grains, respectively, for the refractory 
and volatile cosmic rays. These refractories (mainly Mg, Si, Fe) 
also include C and O because sufficient O is bound in oxides and C 
can condense as graphite. The volatiles are mainly H and He. In the 
CRI model, the cosmic ray source composition is that of the evolving 
ISM, corrected for shock acceleration. The CRS and CRI models lead 
to primary and secondary LiBeB evolution, respectively.

In order to account for the possible rise of [O/Fe] with decreasing 
[Fe/H] (see Figure~1), we suggested that the deposition into star 
forming regions of the ejected Fe is delayed relative to that of O 
because of the incorporation of the refractories of the ejecta into 
fast grains. Note that the bulk of the ejected O is not in grains, 
only the O bound in oxides is refractory and it constitutes the 
cosmic ray source material in the CRS model. The incorporation of 
the bulk of the ejected refractories into fast grains is supported 
by the observation of a broad gamma ray line at 1.809 MeV (Naya et 
al. 1996) resulting from radioactive $^{26}$Al. The best explanation 
for this broadening is the incorporation of the bulk of the 
synthesized Al into fast supernova grains which must still be moving 
at velocities of about 450 km/s some million years (the mean life of 
$^{26}$Al) after the synthesis in supernovae. The results are shown 
in Fig.~1 where we see that with the mixing delays taken into 
account [O/Fe] indeed rises with decreasing [Fe/H] for both the WW95 
and TS models. Our model leads to a prediction for the $\alpha$ 
elements. Specifically, that volatile S should behave as O, i.e. 
that S/Fe should rise, while the abundances of refractory  Mg, Si, 
Ca and Ti relative to Fe should not. This prediction was recently 
confirmed (Takeda et al. 2000). 

Fig.~2 shows the evolution of Be/H as a function of both [Fe/H] and 
[O/H]. Considering the slopes of the data and of the calculated 
curves, we note that even though the overall slope of log(Be/H) vs. 
[Fe/H] is practically unity, while that of log(Be/H) vs. [O/H] is 
significantly steeper (0.96$\pm$0.04 and 1.45$\pm$0.04, 
respectively, Boesgaard et al. 1999b), the CRS model provides a good 
fit to these evolutionary trends, particularly if n$_{\rm H}$ is 
near 0.1. This is a possible value for an early Galactic spherical 
halo of radius 10 kpc containing 10$^{10}$M$_\odot$ of gas. The 
calculated log(Be/H) vs. [Fe/H] is flatter than log(Be/H) vs. [O/H] 
because the delayed deposition of the cosmic ray produced Be, caused 
by the low n$_{\rm H}$, is compensated by the delayed Fe deposition, 
but not compensated by the lack of significant delay for O. We thus 
see that primary evolution of Be vs. both [Fe/H] and [O/H], 
predicted by the CRS model, is possible if the delays due to 
cosmic-ray transport and Fe incorporation into high velocity grains 
are properly taken into account. Thus the steeper evolution of the 
Be abundance vs. [O/H] should not be simply interpreted as secondary 
evolution (Fields et al. 1999), because such evolution, predicted by 
the CRI model, is energetically untenable. We see this from the CRI 
curves in Fig.~2, which severely underpredict the data, independent 
of whether the Be evolution is considered as a function of [Fe/H] or 
[O/H]. 

RR wishes to thank Garik Israelian for pointing out the role of S as 
a test of the theory presented in this poster.

\vskip 2 truecm

\section{\huge Conclusions}

{\huge \noindent 1. The delay of Fe deposition relative to O, due to 
the incorporation of supernova ejecta refractories in high velocity 
grains, can account for the rise of [O/Fe]. The predicted behavior 
of refractory Mg, Si, Ca and Ti, and volatile S, appears to be 
confirmed.} 

{\huge \noindent 2. The delayed deposition of Be, due to cosmic 
ray transport in a relatively low density medium, leads to 
consistency of primary Be evolution as a function of both [Fe/H] and 
[O/H].} 

\begin{figure}[t]
  \begin{center}
    \leavevmode
\epsfxsize=13.cm \epsfbox{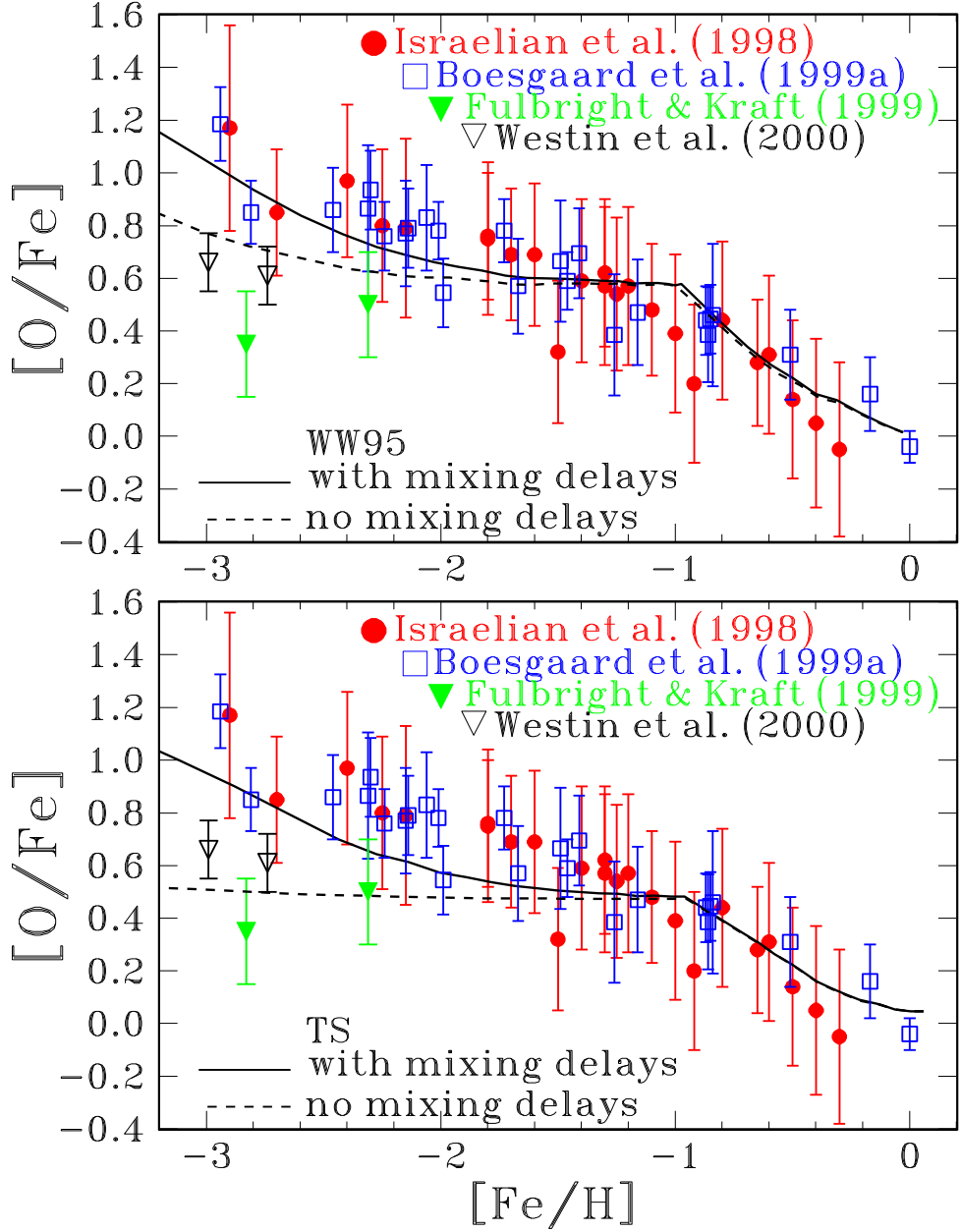}
\end{center}
\caption{The effect on O/Fe evolution of the incorporation of 
refractory Fe into high velocity grains causing its delayed 
deposition into star forming regions.}
\end{figure}

\eject

\begin{figure}[t]
  \begin{center}
    \leavevmode
\epsfxsize=13.cm \epsfbox{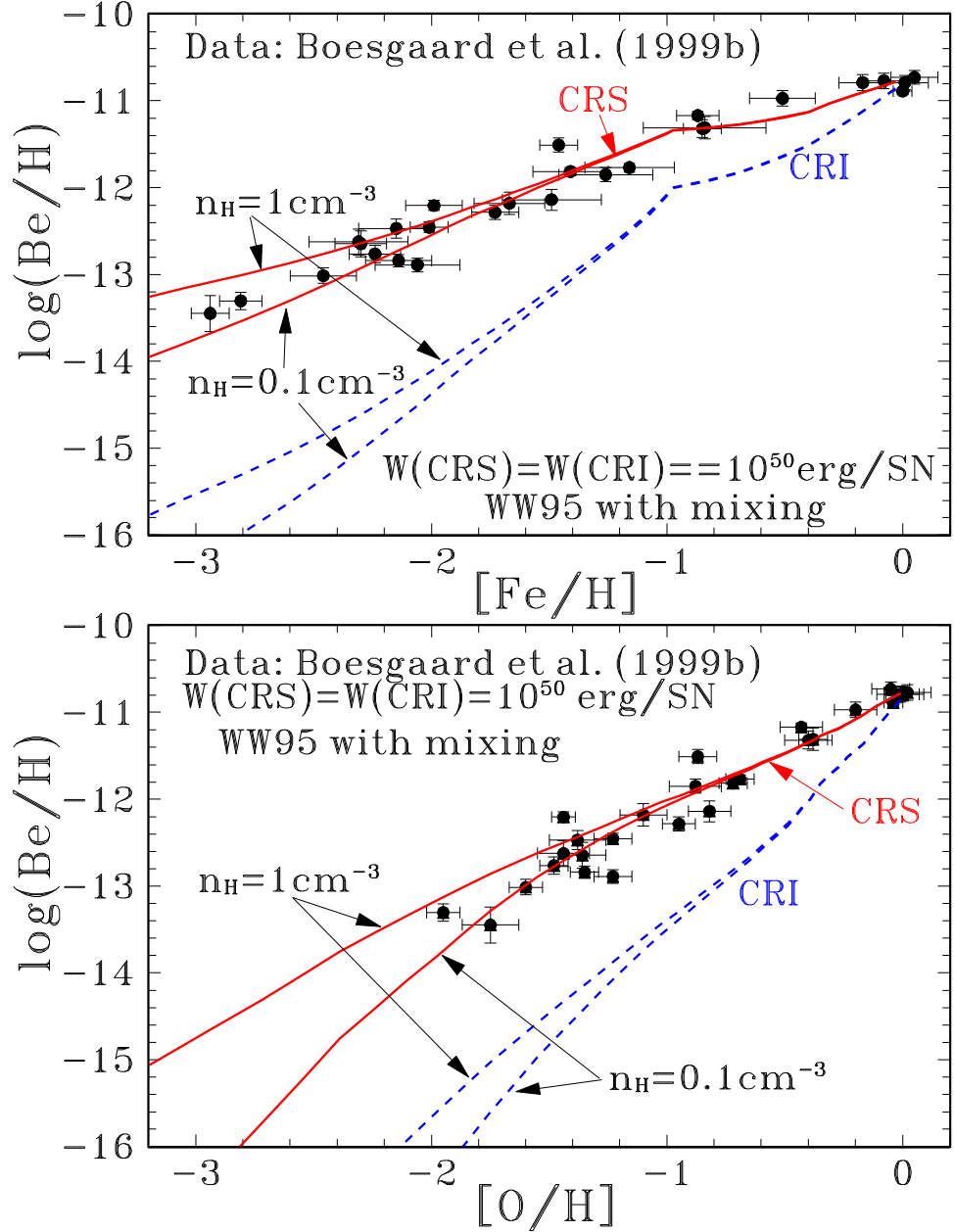}
\end{center}
\caption{Be/H evolution as a function of both Fe/H and O/H. n$_{\rm 
H}$, the density of the ambient ISM, determines the delay of the Be 
deposition due to the cosmic ray transport. CRS and CRI are primary 
and secondary evolution models.}
\end{figure}

\eject

\end{document}